\documentclass[pdflatex,sn-mathphys-ay]{sn-jnl}

\usepackage{graphicx}%
\usepackage{multirow}%
\usepackage{amsmath,amssymb,amsfonts}%
\usepackage{amsthm}%
\usepackage{mathrsfs}%
\usepackage[title]{appendix}%
\usepackage{xcolor}%
\usepackage{textcomp}%
\usepackage{manyfoot}%
\usepackage{booktabs}%
\usepackage{algorithm}%
\usepackage{algorithmicx}%
\usepackage{algpseudocode}%
\usepackage{listings}%
\usepackage[utf8]{inputenc}  

\theoremstyle{thmstyleone}%
\theoremstyle{thmstyletwo}%

\theoremstyle{thmstylethree}%

\begin{document}

\title[Unveiling Complex Territorial Socio-Economic Dynamics: A Statistical Mechanics Approach]{Unveiling Complex Territorial Socio-Economic Dynamics: A Statistical Mechanics Approach}


\author*[]{\fnm{Pierpaolo} \sur{Massoli}}\email{pimassol@istat.it}



\affil*[]{\orgdiv{Directorate for Methodology and Statistical Process Design (DCME)}, \orgname{Italian National Institute of Statistics (ISTAT)}, \orgaddress{\street{Via Cesare Balbo 16}, \city{Rome}, \postcode{00184}, \state{Italy}}}




\abstract{
This study proposes a novel approach based on the Ising model for 
analyzing socio-economic emerging patterns between municipalities  
by investigating the observed configuration of a network 
of selected territorial units which are classified as being central hubs 
or peripheral areas. This is interpreted as being a reference of a system of interacting 
territorial binary units. 
The socio-economic structure of the municipalities is synthesized
into interpretable composite indices, which are further 
aggregated by means of Principal Components Analysis in order to reduce dimensionality and 
construct a univariate external field compatible with the Ising framework. 
Monte Carlo simulations via parallel computing are conducted adopting a 
Simulated Annealing variant of the classic Metropolis-Hastings 
algorithm. This ensures an efficient local exploration of the configuration 
space in the neighbourhood of the reference of the system. 
Model consistency is assessed both in terms of energy stability and 
the likelihood of these configurations. 
The comparison between observed configuration and 
simulated ones is crucial in the analysis of multivariate 
phenomena, concomitantly accounting for territorial interactions. 
Model uncertainty in estimating the probability of each municipality 
being a central hub or peripheral area is quantified by adopting 
the model-agnostic Conformal Prediction framework which yields 
adaptive intervals with guaranteed coverage. 
The innovative use of geographical maps of the prediction 
intervals renders this approach an effective tool.
It combines statistical mechanics, multivariate
analysis and uncertainty quantification, providing a robust and 
interpretable framework for modeling socio-economic territorial 
dynamics, with potential applications in Official Statistics.
}




\maketitle

\section{Introduction}\label{sec1}

The increasing availability of geo-referenced demographic and economic data  
constitutes a challenging opportunity in spatial data analysis. Descriptive  
approaches may prove to be inadequate when managing data complexity if  
territorial structures and socio-economic aspects are combined. Traditional  
spatial-econometric approaches are usually based on autocorrelation and regional  
dependencies, providing important state-of-the-art methodology in applied  
statistics \citep{anselin1988, bivand2013}. These models require a non-trivial  
construction of a pre-defined spatial weight matrix and rely on assumptions of  
linearity and normality of input data, which may not hold in real-world  
applications. As a result, they may fail to capture complex non-linear  
relationships as well as the spatial heterogeneity of socio-economic data,  
resulting in a limited generalization capability.  
Composite indices methodology offers a more flexible approach for analyzing the  
influence of diverse factors related to local units such as regions or  
municipalities, providing a valid alternative to the aforementioned descriptive  
and traditional methods. These widely used methods are adopted in official  
statistics to synthesize multiple dimensions of complex phenomena into a single  
value, enabling territorial comparisons across a set of statistical units for the  
evaluation of multidimensional well-being, deprivation, and social vulnerability.  
A most relevant example is the research work carried out by the Italian BES 
(equitable and sustainable well-being, Italian acronym) Committee for measuring 
multidimensional economic, social, and policy domains \citep{istat2024}. These 
measures support the understanding of complex socio-economic phenomena and 
facilitate effective communication between researchers and policy makers, 
especially when rankings of statistical units are analyzed.
Composite indicators play a central role in the field of social indicators, 
offering essential tools for monitoring inequality, well-being, and regional 
cohesion. They are widely adopted by national statistical institutes and 
international frameworks to evaluate social progress and guide public policy 
across space \citep{noll2004social, greco2019mcda}. These indicators provide a 
multidimensional representation of well-being, going beyond income-based 
measures and enabling spatially informed decision-making.
Despite their widespread use, most composite indices rely on simple additive 
aggregation functions. Although broadly accepted, these methods often assume 
complete compensability across dimensions---meaning that a deficit in one 
dimension can be offset by a surplus in another---which may not hold in 
real-world settings. To address this limitation, a non-compensatory composite 
index has been proposed for both spatial and spatial-temporal comparisons 
\citep{mazziotta2016}. 
A critical limitation remains: composite indices, whether compensatory 
or not, do not capture interactions between statistical units. As such, they may 
prove inadequate for analyzing complex territorial dynamics that emerge from 
interdependencies among local units.
In the field of machine learning, spatial methods are gaining interest in official  
statistics, where predictive modeling of heterogeneous socio-economic data is a  
standard task, as in the case of convolutional neural networks for estimating  
poverty by analyzing satellite imagery combined with socio-economic predictors  
\citep{jean2016}. This integration of spatial statistics with data-driven  
approaches is particularly effective in gaining a deeper insight into social  
inequalities across a given territory.  
In the same context of territorial analysis using machine learning, a novel  
approach based on region-specific boosted classification trees has been proposed  
to evaluate the importance of a set of composite dimensions in relation to an  
observed classification of Italian municipalities as either central hubs or  
peripheral areas \citep{casacci2024}. The results of this approach illustrate the  
potential of using composite indices in machine learning while considering the  
territory as divided into a set of disjoint partitions, without accounting for  
interactions between local units. These limitations suggest that models  
integrating local spatial interactions with external influences to interpret  
observed territorial arrangements of local units may prove to be more robust 
when compared to other classification approaches.  
Graph-based models have recently gained increasing attention for their ability 
to represent spatial systems through topological rather than strictly 
geographical adjacency. In these approaches, spatial units are interpreted 
as nodes in a network, with edges encoding various forms of structural 
or functional similarity rather than physical contiguity. This allows 
for the modeling of spatial interactions in highly heterogeneous 
systems where classical distance-based metrics may be inadequate. 
In human geography and urban planning, graph representations have 
been employed to study socio-spatial organization, infrastructure 
networks, and regional dependencies \citep{liu2022graph, batty2021simulating}. 
These models emphasize the conceptual nature of spatial relationships, offering a flexible 
framework for representing territorial complexity. The approach proposed in this 
study builds on this graph-based perspective by modeling socio-economic interactions between 
municipalities based on shared territorial profiles, independently of 
strict geographical proximity. 
In this perspective, this study proposes the Ising model from Statistical  
Mechanics, a versatile algorithm for describing interactions among particles in  
complex physical systems \citep{ising1925}. 
Recent studies have emphasized the potential of adopting tools from 
statistical physics and network science for modeling socio-economic 
and territorial systems. Approaches based on the Ising model and its 
variants have been successfully applied to understand regional disparities 
\citep{schaefer2023}, structural dependencies in socio-economic resilience 
\citep{duan2022}, and spatial complexity in urban systems \citep{jia2024}. 
These works suggest that integrating spatial interactions within probabilistic 
frameworks can enhance the interpretability and robustness of territorial 
models. This method has also been successfully  
applied to opinion dynamics, biological networks, and socio-economic territorial  
domains \citep{galam1997, durlauf1999}. A key feature of this model consists in  
the effective integration of local interactions with external influences in a  
probabilistic modeling framework.  
In the specific context of spatial analysis, this method may 
describe the asymptotic behavior of a network of interacting local 
units subject to an external field of observed influences, by taking the 
adjacency of the units into account as is the case of municipalities 
with territorial attributes subject to socio-economic influences. 
In such a case the adjacency of local units is interpreted in terms of both 
geographical proximity and territorial similarity.  
Due to the combinatorial complexity of such a network, the exploration of all  
possible configurations of the system is not feasible. Therefore, the system is  
simulated by means of Markov Chain Monte Carlo (MCMC) techniques based on the  
Metropolis-Hastings algorithm, in order to generate configurations of binary local  
units sampled from an unknown probability distribution as the system reaches  
stationarity \citep{metropolis1953, binder1997}.  
This simulation process may be computationally intensive. To reduce the execution  
time and computational burden, a parallel computing approach is mandatory in most  
real-world applications \citep{mccallum2011, weston2017}.  
The randomly generated configurations are used to estimate the probability of each  
local unit being in a particular state, thereby assessing the contribution of  
socio-economic aspects while accounting for territorial similarities in the  
emergence of specific patterns in the estimated classification of municipalities.  
The comparison with the observed classification of municipalities is crucial for  
evaluating the socio-economic influences, which are defined by a set of composite  
indices summarizing social, demographic, and economic characteristics. Principal  
Components Analysis is then applied to further aggregate these indices into a  
single latent structure, referred to as the external field in the Ising model.  
The model is simulated using Simulated Annealing (SA), a variant of the  
aforementioned MCMC approach, aimed at iteratively exploring the configuration  
space by flipping nodes according to local energy changes \citep{aarts1989,  
geman1984}. Simulations are initialized from the observed territorial  
classification, which serves as the reference configuration.  
As reported in the literature, this algorithm is asymptotically convergent to  
a global optimum (if it exists) when a logarithmic cooling schedule is adopted, regardless of the  
initial configuration and initial temperature of the system. This study proposes an  
approach that diverges from this theoretical setting by employing a localized  
stochastic search for more energetically favorable and more likely configurations  
in the neighbourhood of the reference.  
As a consequence, an adaptation of the standard Ising framework is proposed, where  
the updating mechanism is modified in a stochastic gradient descent perspective.  
The reliability of probability estimates is assessed through the Conformal  
Prediction (CP) framework \citep{vovk2005, shafer2008, angelopoulos2021,  
tibshirani2023}. The methods in this framework evaluate distribution-free  
prediction intervals with guaranteed coverage of the true value of the target  
variable of a machine learning model, even in the absence of traditional model  
assumptions \citep{lei2018, burnaev2014}.  
The proposed approach aims to contribute to the field of spatial analysis by  
combining methods from Statistical Physics, Multivariate Analysis, and Conformal  
Prediction. The Ising model framework is innovatively extended for socio-economic  
analysis. The incorporation of Conformal Prediction introduces a crucial step for  
quantifying uncertainty through the construction of adaptive prediction intervals,  
which are projected onto geographical uncertainty maps. These maps represent a  
valuable tool for detecting areas of limited model reliability in policy-driven  
territorial analysis, as highlighted by the results from the real-world case study  
presented in this article.

\section{Theoretical background}\label{sec2}

In order to introduce the basic aspects of the proposed approach to 
the reader, some notions regarding the Ising model as well as the 
composite indices method being adopted and Conformal Prediction 
framework are reported in this section. 

\subsection{The Ising model}\label{sec2_1}

The Ising model, formerly proposed for studying ferromagnetism in materials 
in a Statistical Mechanics context, is also applied to more general systems  
where the effect of local interactions is important. 
It is a discrete system constituted by a network of $N$ nodes, 
each assuming a binary state $s_i \in \{-1, +1\}$ ($i=1, 2, \ldots, N$) called \emph{spin}. 
A configuration of the system is denoted by $\mathbf{s}=\{s_1, s_2, \ldots, s_N\}$ which is 
an element of the set $\mathbb S$ of all possible 
configurations. A straightforward mathematical representation of the 
network is the undirected graph.  
The energy (\emph{Hamiltonian}) of a configuration is defined as follows
\begin{equation}\label{eq:hamilt}
H(\mathbf{s}) = -\frac{1}{2}\:\sum_{i, j}^{N} J_{ij}s_is_j - \sum_{i}^{N} h_i s_i
\end{equation}
where the element $J_{ij}$ of the symmetric matrix $\mathbf J$ 
defines the interaction between nodes $i$ and $j$ while $h_i$ is the 
external field acting on  node $i$. By representing the network as 
being a graph, $\mathbf J$ may be interpreted as the weighted adjacency matrix, where $J_{ij} > 0$ 
indicates the pair $(i, j)$ of connected nodes. 
As the system reaches thermodynamic equilibrium, the probability 
of observing a given configuration $\mathbf{s}$ is defined by means of 
the Boltzmann distribution:
\begin{equation}\label{eq:ll}
P(\mathbf{s}) = \frac{1}{Z} e^{-\frac{H(\mathbf{s})}{T}},
\end{equation}
where $T$ is the \emph{temperature} of the system and 
$Z$ is the \emph{partition function} defined as follows:
\begin{equation}
Z = \sum_{\mathbf{s} \in \mathbb{S}} \exp(- H(\mathbf{s})).
\end{equation}
This distribution assigns a higher probability to configurations  
related to lower energy values, depending on the temperature of the 
system. This fixed parameter influences the physical behavior of 
the system so that when $T$ is high the system is chaotic while lower 
the temperatures imply configurations pertaining to minimum values 
of energy. The general Hamiltonian function is non-convex with 
a highly irregular surface, suggesting the Simulated 
Annealing (SA) variant of the classic Metropolis-Hastings algorithm 
which is usually adopted in the standard Ising framework. 
The interaction matrix $\textbf{J}$ results to be \emph{indefinite}\footnote{
In Linear Algebra, a quadratic form $q(\mathbf{x}, \mathbf{y})=\mathbf{y}^{T}A\mathbf{x}$  with $\mathbf{x}, \mathbf{y} \in \mathbb{R}^n \setminus \{ \mathbf{0} $ is 
indefinite if the symmetric matrix $\mathbf A$ admits both 
positive and negative eigenvalues so that $q(\mathbf{x}, \mathbf{y})>0$ or $q(\mathbf{x}, \mathbf{y})<0$.} 
in many practical cases,  implying the existence of multiple local minima 
as well as saddle points of the Hamiltonian function. 
As a result, standard MCMC method may get stuck in sub-optimal 
configurations which are best avoided by a SA strategy.
This variant requires a time-dependent temperature cooling schedule 
$T=f(t)$, decreasing over time, and updates the solution in accordance 
with an acceptance criterion. The \emph{acceptance probability} of a proposed 
spin configuration at time $t$ is defined as follows:
\begin{equation}
	\alpha = \min\Big\{1, \exp\left(-\frac{\Delta H}{T}\right)\Big\}
\end{equation}
where $\Delta H$ is the (local) change in energy subsequent to the 
flipping of the spin pertaining to a node selected at random. 
This node selection is iteratively repeated during the MCMC 
simulation in order to span the entire network, yielding an 
approximate estimate of the total energy variation in a Stochastic 
Gradient Descent perspective.
As the system cools down it converges to a minimum of the energy 
gradually by escaping local minima. 
In real systems when the interaction structure  
is a large weighted graph, the  
partition function $Z$ becomes computationally intractable, 
hindering a direct computation of probabilities as is indicated 
in Equation~\ref{eq:ll}. As a consequence, MCMC approaches are 
adopted in order to generate system configurations sampled from 
the aforementioned probability distribution. 

\subsection{Methods for creating composite indices }\label{sec2_2}

Composite indices are widely adopted in the social sciences when  
the synthesis of multi-dimensional information into a single value  
is required for taking the overall performance of statistical units 
into account with respect to specific phenomena. 
Composite indices facilitate the comparison between  
different statistical units. Compensatory indices provide 
a compensation of low values in a base indicator with high values 
in another if the assumption that they are substitutable is valid. 
Non-compensatory composite indices are suitable when all input 
dimensions are essential. 

\paragraph{A popular non-compensatory composite index} 

A well-known method in the literature for constructing 
composite indices is the \textit{Mazziotta-Pareto Index} 
(MPI), a non-compensatory method based on a standardization 
of base indicators in \textit{z-scores} for a subsequent aggregation 
by penalizing unbalanced unit profiles. 
Suppose an input dataset $\textbf{X}=\{x_{ij}\}$ containing 
$j=1,2, \ldots, M$ base indicators pertaining to 
$i=1,2, \ldots, N$ statistical units, the MPI composite index 
requires the standardization of every value in the dataset 
as follows
\begin{equation}
	x_{ij}^{s} = 10 \cdot pol_{j} \cdot \left( \frac{x_{ij} - \mu_j}{\sigma_j} \right) + 100
\end{equation}
where $x_{ij}$ is the original $i$-th value of the $j$-th base indicator, 
$pol_j$ its polarity (equal to $+1$ or $-1$), $\mu_j$ its mean, and 
$\sigma_j$ its standard deviation. 
The MPI is subsequently computed as follows:  
\begin{equation}
MPI_i = M_i \pm S_i \cdot \frac{S_i}{M_i},
\end{equation}
where $M_i$ and $S_i$ are the mean and standard deviation  
of the standardized profile $i$, respectively. The  
sign $\pm$ depends on whether the phenomenon being under consideration 
is  positive or negative. A higher dispersion among the input base 
indicators affects the final score of the composite index, reflecting  
a penalization for unbalanced units. 
In order to create a meaningful index in relation to 
a specific aspect under examination, all its base indicators 
have to be related to the same as well. 
The polarity of each indicator may assume opposite 
signs insofar as they all have the same direction. 
This crucial aspect of the composite index construction has to be 
addressed by researchers before proceeding to the construction. 

\paragraph{An effective compensatory composite index}

Another well-known method in the literature for synthesizing information is 
\emph{Principal Components Analysis} (PCA), a dimensionality reduction 
technique that transforms the original correlated variables into a set of 
uncorrelated components. The first few principal components typically 
maintain most of the variance in the data, allowing for a simplified yet informative 
representation.
In this context, the principal components are used to construct a weighted 
composite index, where weights are derived from the explained variance of 
each component. The dataset $\textbf{X}$ is decomposed into its principal 
components $\textbf{pc}_1, \textbf{pc}_2, \ldots, \textbf{pc}_M$, 
which are aggregated as follows:
\begin{equation}
	\textbf{C}_\text{PPCA} = \lambda_{1}\textbf{pc}_{1}+\lambda_{2}\textbf{pc}_{2}+\ldots+\lambda_{M}\textbf{pc}_{M}
\end{equation}
where $\lambda_i$ indicates the $i$-th normalized eigenvalue of the PCA 
decomposition of the input dataset\footnote{The weights $\lambda_i$ used 
in the construction of the composite index are normalized so that their 
sum is equal to one.}. Each $\lambda_i$ represents the proportion 
of total variance explained by the $i$-th principal component. 
This approach ensures that the composite index reflects 
the dominant structure of variability in the original data

\subsection{A quick glance at Conformal Prediction}\label{sec2_5}

Conformal Prediction (CP) is a versatile framework for quantifying 
uncertainty in a machine learning model $y=f(\textbf{x})$, yielding guaranteed coverage 
relating to its prediction intervals without requiring assumptions about data 
distribution. This framework is particularly robust insofar as model 
reliability is a critical concept as far as practical applications are 
concerned. Prediction intervals are evaluated by means of 
a calibration set in order to compute \emph{non-conformity scores}. 
In this context these latter are calculated as standardized residuals 
defined as follows:
\begin{equation}\label{eq:CP.scores}
	s_{i} = \frac{|y_i - \hat{y}_i|}{u(\textbf{x})},
\end{equation}
where $y_i$ and $\hat{y}_i$ are respectively the observed (true) and 
predicted values and $u(\textbf{x})$ represents a measure of the uncertainty 
related to the input vector $\textbf{x}$ computed by using data belonging to the 
calibration set. 
The prediction intervals for every observation $\bf x$ 
belonging to the test set are evaluated as follows:
\begin{equation}\label{eq:CP.PI}
	\mathcal{C}({\bf x})=[\hat{y}_i - \hat{q}\sigma, \hat{y}_i + \hat{q}\sigma],
\end{equation}
where $\hat{q}$ is the $\lceil(1-\alpha)(n+1)\rceil/n$ quantile with 
$n$ equal to the number of observations in the calibration dataset and 
$\alpha$ is the user-defined level of accuracy. 
This split-conformal approach requires that the dataset is partitioned 
into disjoint calibration and test sets. 
The CP framework requires a dataset of $n$ observations 
$\{(X={\bf x}_{i}, Y=y_{i})\}$ ($i=1, 2, \ldots, n$) 
with features $X \in \mathbb{R}^d$ and response $y \in \mathbb{R}$, 
the prediction interval $\mathcal{C}(X)=[L(X), U(X)]$ covers the 
true value with probability:
\begin{equation}\label{eq:CP.covprob} 
	\Pr(y \in \mathcal{C}(X)) \geq 1 - \alpha,
\end{equation}
In order to further evaluate the properties of prediction intervals 
whilst ensuring a pre-fixed coverage the \emph{Mean Interval Width} (MIW) 
is used. It is defined as:
\begin{equation} 
	MIW = \frac{1}{n} \sum_{i=1}^{n} (U_i - L_i) 
\end{equation}
where $U_i$ and $L_i$ are the upper and lower bounds of the prediction interval 
for observation $i$ in a test set containing $n$ observations. 
In order to normalize this measure relative to the scale of the target 
variable, the \emph{Relative Interval Width} (RIW):
\begin{equation} 
	RIW = \frac{\text{MIW}}{y_{\max} - y_{\min}} 
\end{equation}
where $y_{\max}$ and $y_{\min}$ denote the maximum and minimum 
observed values of the target variable. A lower MIW indicates narrower 
intervals, while a lower RIW facilitates comparisons across different 
datasets or models by adjusting for data scale variations.
These measures yield a precise evaluation of the adaptivity 
of the prediction intervals.

\section{Proposed approach}\label{sec3}

The proposed approach is grounded on the application of the Ising model to 
the analysis of binary territorial classifications, integrating local 
interactions and external influences within a probabilistic framework. This 
methodology is designed to be general and adaptable to any spatial context 
in which local units are described by a set of socio-economic characteristics 
and territorial features, and a binary classification is available.
Each local unit is characterized by a set of base indicators capturing 
demographic, educational, economic, and occupational aspects, as well as 
territorial attractiveness and dynamism. These indicators are aggregated 
into composite indices by means of the Mazziotta-Pareto methodology. 
Dimensionality is reduced through Principal Components Analysis (PCA), 
which is used to construct the external field of the model, as illustrated 
in the case study which follows.
Territorial similarities are incorporated through the construction of a 
weighted undirected graph where nodes represent local units. The edges 
reflect proximity or similarity based on features such as altitude, surface 
area, population size, degree of urbanization and coastal proximity. The 
resulting interaction matrix defines the topology of the Ising network. 
Model simulation is performed using a variant of the Metropolis-Hastings 
algorithm based on Simulated Annealing. The observed configuration is 
considered as a starting point of the algorithm which explores the 
space of possible configurations by changing the state of one node at a time. 
The proposed Simulated Annealing scheme, guided by a rapidly 
decreasing temperature schedule, performs a localized stochastic search 
around this reference. Rather than seeking a global minimum, the model 
aims to identify nearby, energetically favorable configurations, thus 
reducing sensitivity to the initial state over long simulations and 
supporting the interpretative focus of the approach.
The probability of each local unit being in a given state is estimated 
from the marginal distribution obtained by generating multiple 
configurations.
In order to quantify the uncertainty of these estimates, the Conformal 
Prediction framework is employed, deriving distribution-free prediction 
intervals with guaranteed coverage under minimal assumptions. 
The uncertainty quantified by the model is primarily epistemic, 
stemming from structural assumptions and aggregation processes, 
rather than data randomness. 
As a consequence, the absence of a sensitivity 
analysis on the MCMC parameters is compensated by adopting the 
Conformal Prediction  framework as the uncertainty estimation accounts for 
the total variability induced by the model structure.
This framework provides a robust methodology for analyzing territorial 
configurations as emergent phenomena shaped by endogenous interactions 
and exogenous socio-economic pressures, with particular applicability to 
complex heterogeneous systems.
The estimated marginal distribution is therefore not an end in itself, 
but a means to generate plausible territorial classifications of the 
municipalities in terms of $-1/+1$ spin values. These simulated 
configurations represent alternative arrangements that are coherent 
with the model assumptions and can be compared to the observed classification. 
This comparison enables a deeper investigation of the socio-economic dimensions that 
drive the emergence, deviation, or ambiguity of spatial patterns across the territory.

\section{A real-world application to Italian municipalities}\label{sec4}

In order to illustrate the potential of the proposed method, a case study 
regarding real-world data was analyzed. Data were taken from the 
statistical register \textit{ARCHIMEDE} of the Italian National 
Institute of Statistics (ISTAT). The register integrates diverse 
administrative sources, providing a detailed description of all 
Italian municipalities.
The main objective of this case study is not to claim superiority of the 
proposed approach over existing classification methods, rather to 
investigate whether a statistical mechanics-based model can reproduce 
and explain spatial patterns emerging from territorial interactions. 
Nonetheless, a comparative benchmark with standard models is presented 
in Section~\ref{sec:benchmark}, as a final step to assess classification 
performance from a predictive perspective.

\subsection{The input dataset}\label{sec4_0}

The input dataset was extracted from the aforementioned register by 
selecting a subset of relevant socio-economic base indicators. 
This dataset is made up of a total of $7908$ municipalities with profiles 
with no missing data, each being characterized by a set of socio-demographic, 
economic and territorial variables as well as geographic attributes. 
Each municipality is classified as being a  \emph{central hub} 
or \emph{peripheral area} based on exogenous criteria.
As the available territory is rather large and non-homogeneous, the 
results may be difficult to interpret and the model simulations may 
be computationally intensive; the only territory which was taken 
into consideration was the central macro-region made up by the 
following regions: \emph{Lazio}, \emph{Marche}, \emph{Tuscany} and 
\emph{Umbria}. As a result, the total number of municipalities was reduced to $966$.
This case study analyzed the configurations of municipalities 
which were generated by using the Ising model in accordance with the 
observed one which is the reference configuration. The objective was to investigate 
the impact of socio-economic aspects on the probability of municipalities 
being central hubs, having taken territorial structures into account. 
The list of features is reported in Table~\ref{tab:mpi_detailed}. 
\begin{table}[h]
\centering
\caption{Description of Composite Indices}
\label{tab:mpi_detailed}
\tiny
\begin{tabular}{p{0.1\linewidth}lp{0.25\linewidth}r||}
\toprule
\textit{Composite Index} & \textit{Base Indicator} & \textit{Description} & \textit{Polarity} \\
\midrule
\multirow{4}{*}{\begin{minipage}[t]{\linewidth}MPI1\\(Demographic Structure)\end{minipage}} 
& PERC\_ANZIANI & \% of elderly (65+) in the population & -1  \\
& PERC\_GIOVANI & \% of young people (<15) in the population & +1  \\
& PERC\_FAMIGLIE\_MINORI & \% of households with minors & +1  \\
& PERC\_FAM\_UNIPERSONALI\_ANZIANI & \% of single-member elderly households & -1  \\
\midrule

\multirow{3}{*}{\begin{minipage}[t]{\linewidth}MPI2 \\ (Cultural Level)\end{minipage}} 
& PERC\_NEET & \% of NEETs (not in education, employment or training) & -1  \\
& PERC\_LAUREATI & \% of university graduates & +1  \\
& PERC\_DIPLOMATI & \% of high school graduates (25-64 years old) & +1  \\
\midrule

\multirow{2}{*}{\begin{minipage}[t]{\linewidth}MPI3 \\ (Economic Well-being)\end{minipage}} 
& REDDITO\_MEDIANO\_EQUIVALENTE & Median equivalent income (log-transformed) & +1  \\
& PERC\_WORKINGPOOR & \% of workers earning less than 60\% of the median wage & -1  \\
\midrule

\multirow{3}{*}{\begin{minipage}[t]{\linewidth}MPI4 \\ (Employment Level)\end{minipage}} 
& PERC\_PRECARI & \% of precarious workers (October snapshot) & -1  \\
& PERC\_OCCUPATI & \% of employed individuals (20-64 years old) & +1  \\
& PERC\_FAM\_BASSA\_INTLAV & \% of households with low work intensity & -1  \\
\midrule

\multirow{3}{*}{\begin{minipage}[t]{\linewidth}MPI5 \\ (Territorial Attractiveness)\end{minipage}} 
& I\_ATTRAZIONE & Attraction index & +1  \\
& I\_AUTOCONTENIMENTO & Self-containment index & +1  \\
& I\_COESISTENZA & Coexistence index & +1  \\
\midrule

\multirow{4}{*}{\begin{minipage}[t]{\linewidth}MPI6 \\ (Population Dynamism)\end{minipage}} 
& STA & Static individuals (no signs of work/study activity) & -1  \\
& D\_INT & Internal movers (within the same municipality) & -1  \\
& D\_EST\_USCITA & External movers leaving the municipality & +1  \\
& D\_EST\_ENTRATA & External movers entering the municipality & +1  \\
\bottomrule
\end{tabular}
\normalsize	
\end{table}
The polarities of the base indicators are set up in order to 
calculate coherent values of composite indices in the real-world 
scenario being considered. 
These indices constitute the univariate external 
field of the Ising model as is reported in the following. 
Territorial attributes of municipalities are used for modelling 
interactions between units; all those ones sharing the same territorial 
profile are connected with a weight equal to $1$ even though they 
may be not neighbouring municipalities. The territorial attributes 
are reported in Table~\ref{tab_2}. 
\begin{table}[h]
\centering
\caption{Territorial attributes of municipalities}
\label{tab_2}
\tiny
\begin{tabular}{p{0.25\linewidth} l p{0.6\linewidth}}
\toprule
\textit{Attribute} & \textit{Item} & \textit{Description} \\
\midrule

\multirow{3}{*}{\begin{minipage}[t]{\linewidth}ALT \\ (Altitude of the centre)\end{minipage}} 
& 1 & Lowland (below 300 meters) \\
& 2 & Hill (between 300 and 600 meters) \\
& 3 & Mountain (above 600 meters) \\
\midrule

\multirow{3}{*}{\begin{minipage}[t]{\linewidth}POP \\ (Resident population Size)\end{minipage}} 
& 1 & Small (fewer than 5,000 inhabitants) \\
& 2 & Medium (between 5,000 and 50,000 inhabitants) \\
& 3 & Large (more than 50,000 inhabitants) \\
\midrule

\multirow{3}{*}{\begin{minipage}[t]{\linewidth}SUP \\ (Surface Area)\end{minipage}} 
& 1 & Small (less than 15 km\textsuperscript{2}) \\
& 2 & Medium (between 15 and 100 km\textsuperscript{2}) \\
& 3 & Large (more than 100 km\textsuperscript{2}).\\
\midrule

\multirow{2}{*}{\begin{minipage}[t]{\linewidth}CLITO \\ (Coastal Location)\end{minipage}} 
& 0 & Non-coastal \\
& 1 & Coastal \\
\midrule

\multirow{3}{*}{\begin{minipage}[t]{\linewidth}DEGURB \\ (Urbanization Level)\end{minipage}} 
& 1 & City / Densely populated areas \\
& 2 & Towns and suburbs / Intermediate density areas \\
& 3 & Rural areas / Sparsely populated areas \\
\bottomrule
\end{tabular}
\end{table}
As a result, the network becomes a simple undirected graph in 
which each node identifies a municipality of the territory being 
considered. The interaction matrix $\textbf{J}$ of Equation~\ref{eq:hamilt} reduces 
therefore to the adjacency matrix with ${0, 1}$ elements. 
The observed configuration is considered as being the reference 
configuration of spins $\{-1, +1\}$ for the MCMC simulations.
The base indicators listed in Table~\ref{tab:mpi_detailed} were selected 
solely on the basis of their availability. Even though the addition of 
variables such as transport infrastructure coverage, commuter flows 
and economic networks may further refine the modeling of territorial 
interactions for yielding more coherent configurations, a deeper 
investigation into the optimal selection of base indicators is beyond 
the scope of this study. 
It is important to point out that the interaction matrix was constructed 
by connecting municipalities which share similar structural and 
territorial attributes regardless of their geographical 
contiguity. As a result, this matrix describes a \emph{conceptual network}, 
which is more suitable to the proposed approach in this study. 
as opposed to geographical adjacency, which may connect highly 
dissimilar municipalities, this formulation promotes 
interactions among units with comparable profiles, supporting 
a more interpretable modeling of territorial structures.

\subsection{External field of the Ising model}\label{sec4_2}

Before applying the PCA as is described in Section~\ref{sec2} in order  
to create the external field,  the \emph{Pearson correlation} matrix 
between the input composite indices ($MPI.1$-$MPI.6$) was computed. 
\begin{table}[h]
\centering
\caption{Correlation of the composite indices}
\label{tab:corr_mpi}
\tiny
\begin{tabular}{lcccccc}
\toprule
       & MPI.1 & MPI.2 & MPI.3 & MPI.4 & MPI.5 & MPI.6 \\
\midrule
MPI.1 & 1.000  & 0.255 & 0.170 & 0.172 & 0.177 & 1.000 \\
MPI.2 & 0.255  & 1.000 & 0.441 & 0.428 & 0.240 & 0.255 \\
MPI.3 & 0.170  & 0.441 & 1.000 & 0.893 & 0.046 & 0.170 \\
MPI.4 & 0.172  & 0.428 & 0.893 & 1.000 & 0.069 & 0.172 \\
MPI.5 & 0.177  & 0.240 & 0.046 & 0.069 & 1.000 & 0.177 \\
MPI.6 & 1.000  & 0.255 & 0.170 & 0.172 & 0.177 & 1.000 \\
\bottomrule
\end{tabular}
\end{table}
This matrix reveals that the correlation between $MPI.3$ (economic well-being) 
and $MPI.4$ (employment) assumes a value equal to $0.893$ 
as well as the correlation between $MPI.1$ and  $MPI.6$ equal 
to $1$ highlight linear dependency between them while $MPI.5$ 
(attractiveness of the municipality) reflects a weakly correlated 
input dimension in relation to the other ones, revealing redundant 
information which motivates the adoption of the dimensionality 
reduction approach by means of PCA. 
As a consequence, the composite indices in the input dataset 
$\textbf{X}$ were transformed into uncorrelated dimensions so that 
$\textbf{X}=\{\textbf{pc}_1, \textbf{pc}_2, \ldots, \textbf{pc}_6\}$ 
where the vector $\textbf{pc}_i$ indicates the $i$-th principal 
component. The objective is to create a \emph{super-composite index} 
which is not affected by input data dependencies yet maintaining the 
maximum value of the explained variance of the same. 
On the basis of this reasoning the external field was defined 
as follows:
\begin{equation}\label{eq:ext_field}
	\textbf{h} = \lambda_{1}\textbf{pc}_{1} + \lambda_{2}\textbf{pc}_{2} + \ldots + \lambda_{6}\textbf{pc}_{6}
\end{equation}
where $\textbf{h} = \{h_1, h_2, \ldots, h_n\}$ is the array 
of the values of the resulting external field pertaining to 
$n$ municipalities which constitute the territorial network 
of the model. The weights $\lambda_i$ being used to construct 
the external field are reported in Table~\ref{tab:pca_summary}.   
\begin{table}[h]
\centering
\caption{Principal Components Analysis summary}
\label{tab:pca_summary}
\begin{tabular}{lcccccc}
\toprule
\textit{Component} & \textit{PC1} & \textit{PC2} & \textit{PC3} & \textit{PC4} & \textit{PC5} & \textit{PC6} \\
\midrule
Standard deviation & 1.7327 & 1.1968 & 0.9416 & 0.6925 & 0.4461 & 0.0000 \\
Proportion of Variance ($\lambda_i$) & 0.5004 & 0.2387 & 0.1478 & 0.0799 & 0.0332 & 0.0000 \\
Cumulative Proportion & 0.5004 & 0.7391 & 0.8869 & 0.9668 & 1.0000 & 1.0000 \\
\bottomrule
\end{tabular}
\end{table}
The number of variables in the input dataset $\textbf{X}$ may be further 
reduced due to the fact that the first three principal components 
are sufficient to explain the $88.69\%$ of the variance of 
the input data. 
The external field is not required to be interpretable as it 
is related to the MPI indices which are synthetized via PCA, performing 
an efficient dimensionality reduction while maintaining relevant 
information.  
The aggregation of the input composite indices by means if the PCA 
 is motivated by the requirement of extracting 
a latent structure which preserves the most of variance while removing 
collinearity among the input data. 
The PCA-based super-index offers a valid synthesis of the input 
composite indices, yielding a continuous univariate variable 
which constitutes the bridge between the Ising model and the multivariate 
socio-economic aspects under investigation.
Despite the complete collinearity of two input variables, all 
six principal components are maintained in the construction of the 
external field. The component related to zero variance does not 
contribute to the weighted sum, as its corresponding eigenvalue 
is zero. As a result, the formulation in Equation~\ref{eq:ext_field} 
remains numerically stable while preserving the full dimensional structure 
of the data.

\subsection{MCMC simulations of the model}\label{sec4_3}

The marginal probability of each local unit being in the central hub 
state is estimated through Monte Carlo simulations. The simulation 
starts from the observed territorial configuration and proceeds by 
flipping spins according to local energy changes.
A total of $N_{\text{iter}}$ iterations is performed, with temperature 
initialized at $T_0 = 100$ and decreasing according to a hyperbolic 
schedule $T(t) = T_0 / t$. This allows the system to rapidly concentrate 
around energetically favorable configurations. At each iteration, a 
local unit is randomly selected, and its spin is flipped with a 
probability depending on the energy variation and current temperature. 
The empirical marginal probability for each unit is computed as the 
frequency of state $+1$ across sampled configurations, representing the 
estimated likelihood of being classified as a central hub.
Since the simulation aims to explore configurations near the observed 
one rather than achieve global convergence, this setup is effective in 
detecting plausible structural alternatives driven by the model dynamics.
The MCMC algorithm requires a large number of iterations 
$t = 1, 2, \ldots, N_\text{iter}$ to produce stable results, making a 
parallel computing approach often necessary. In the Simulated Annealing 
\emph{meta-heuristic} framework, the probability of accepting 
energetically unfavorable configurations ($\Delta H > 0$) depends on 
the cooling profile. A faster rate reduces the probability of such 
acceptances, directing the process towards lower-energy states 
($\Delta H < 0$).
It is well-established in the literature that this optimization 
technique converges asymptotically to an optimum independently of the 
initial configuration and temperature when a logarithmic schedule is 
used.
In this study, the initial configuration is fixed rather than selected 
at random, while temperature is initialized at $T_0 = 100$, a value 
that supports broad initial exploration and gradual focus on the 
reference configuration.
Empirical validation confirmed that varying $T_0$ produces consistent 
estimates. The purpose is not exhaustive global exploration, rather 
to identify energetically plausible configurations in the neighbourhood 
of the observed one. A slower logarithmic schedule may induce unwanted 
deviations from this reference.
In order to achieve this aim, a hyperbolic cooling function $T(t) = T_0 / t$ is 
adopted to accelerate convergence around the observed configuration. 
The setting of this schedule thus reflects the preference for local rather 
than global exploration. 
Each sampled configuration generated by the Ising model corresponds to a 
categorical classification of the entire set of municipalities, where each unit 
is assigned a binary state $s_i \in \{-1,+1\}$ indicating peripheral area or 
central hub, respectively. As a result, the simulation process yields a 
collection of alternative territorial classifications consistent with the 
interaction network and the external field. The empirical marginal probability 
that a given municipality is classified as a central hub is then estimated as 
the relative frequency of state $+1$ across all sampled configurations after 
the burn-in phase. 
This approach assesses model-driven deviations from the observed 
classification rather than seeking a single global minimum.
The simulation algorithm iteratively selects one node at random and 
flips its spin based on the energy change and current temperature, as 
detailed in Section~\ref{sec2_1}. This procedure generates a sequence 
of configurations used to estimate marginal probabilities.
The empirical estimation corresponds to the proportion of sampled 
configurations in which each unit assumes state $+1$, indicating 
classification as a central hub.
Given the rationale above, the absence of a formal sensitivity 
analysis is justified. The uncertainty due to simulation is already 
incorporated into the estimated marginal probabilities, making a 
Conformal Prediction approach a suitable alternative.
The energy and likelihood of sampled configurations are evaluated as 
described in the following section.

\subsection{Energy and likelihood of the simulated configurations}\label{sec4_4}

The coherence of the model was assessed after having estimated 
the marginal probability distribution of each 
municipality, the same is used to generate $N$ new spin configurations 
in order to evaluate the ratio between energy $H$ of each 
$\textbf{s}$ and the reference configuration $H_{ref}$ (see Equation~\ref{eq:hamilt}). 
Energy ratios having most of the values below $1$ indicate a 
reliable and robust system. 
On the basis of this reasoning, the ratio between the 
log-likelihood of each simulated configuration and the reference one 
indicates configurations which are more compatible with the territorial constraints as well as the external field . According 
to Equation~\ref{eq:ll}, the 
ratio is defined as follows:
\begin{equation}\label{eq:ll_ratio}
\text{log} \Big(\frac{P(\mathbf{s})}{P(\mathbf{s}_{ref})}\Big) \propto -\Delta H
\end{equation}
where $\Delta H = H - H_{ref}$ indicates the difference between the hamiltonian of a generic configuration and the reference one. This formulation avoids the 
evaluation of the partition function which is intractable 
in real-world systems\footnote{The log-likelihood ratio is proportional to the energy variation up to 
a scaling factor given by the temperature $T$ of the system which 
is considered as being at a temperature assumed when computing the 
ratio for comparing the simulated configurations to the 
reference.}. 
These measurements are proposed for the selection 
of a set of configurations used for estimating the 
probability distribution in order to investigate the misclassified 
municipalities. 
A further analysis of the spectrum of the interaction matrix 
$\textbf J$ includes the presence of both positive and negative 
eigenvalues, confirming a non-convex Hamiltonian which may exhibit 
multiple local minima\footnote{The determinant of $\textbf J$ is negative resulting equal to 
$\det(\textbf{J}) = -6.24 \times 10^{35}$.}.

\subsection{Uncertainty quantification of the model}\label{sec4_5}

It is important to remind the reader that, 
the probability that each municipality is a central hub is estimated 
as being the empirical mean of its spin value across a sample of 
configurations generated by the model at stationarity. 
As a consequence, the generation of a pre-fixed number of samples of spin configurations is required in 
order to have as many independent marginal probability estimations as the 
number of samples in which 
each sampled configuration 
represents a plausible territorial classification.
Uncertainty of the model in producing these estimates is effectively 
evaluated by determining the prediction intervals of the probabilities 
of being a central hub for each municipality in the network. 
To be more precise, subsequent to the evaluation of the likelihood 
of the sampled configurations as is described in the previous section, $K$ 
matrices of $N$ spin configurations are generated in order to get 
$K$ independent estimations of the aforementioned marginal probability. 
The prediction intervals of the estimated probability for each 
municipality are evaluated according to the Conformal Prediction 
approach reported in Section~\ref{sec2} where the non-conformity 
scores are calculated as described by Equation \ref{eq:CP.scores} 
where the true value of the probability for the $i$-th municipality 
is denoted by $y_i$ while $ \hat{y}_i$ is its $k$-th probability estimate 
($k=1,2, \ldots, K$) . In this study the uncertainty $u(\textbf{x})$ is 
considered equal to the standard deviation $\sigma(\textbf{x})$ of the $K$ 
values of each municipality. 
The idea behind the construction of prediction intervals is that the analysis 
of the coverage as well as the adaptivity of the intervals provides useful insights on the model accuracy in relation to the external socio-economic information spread on the territory. 
Another notable contribution of the Conformal Prediction framework 
adopted in this study is the innovative generation of \emph{uncertainty maps} for highlighting areas where model estimates diverge from the observed 
configuration.
The robustness of the estimated marginal probabilities is empirically 
assessed through the Conformal Prediction framework, instead of 
performing a traditional sensitivity analysis on MCMC parameters 
(i.e. initial temperature and number of iterations). The prediction 
intervals quantify the uncertainty arising from both the simulation 
process and the model structure, thus avoiding the traditional 
sensitivity analysis of the MCMC simulations.

\section{Results}

In order to provide a robust estimation of the aforementioned marginal 
probability, the MCMC simulation of the Ising model was performed by 
means of a total of $600000$ iterations of which 
the first $60000$ ($10\%$) were discarded as they were generated 
during the \emph{burn-in} step of the simulations to ensure a stationary sampling process. 
The remaining $90\%$ of the configurations 
were used to estimate the probability. Simulations were executed in a \emph{parallel 
computing} context on a machine equipped with $8$ logical cores 
(including \emph{hyperthreading}). Six cores 
were utilized for computation, resulting in a total execution time of 
approximately $1.19$~\textit{min}.
The simulations were carried out locally on a laptop featuring an 
11th generation \emph{Intel Core i5-1145G7} processor with $4$ physical 
cores and $8$ threads, and a base frequency of $2.60$~\textit{GHz}. 
The operating system was a $64$-bit version of \emph{Microsoft 
Windows}. All computations were performed locally, without relying on 
cloud-based or distributed computing resources. \\

The selection of MCMC simulation parameters was not the result 
of a sensitivity analysis; parameters were set up on the basis of 
\textit{load-balancing} considerations, rendering an efficient 
use of the six cores available when performing parallel computing. 
The number of iterations was set up in order to allocate a sufficient 
large number of the same to the aforementioned cores. 
\begin{figure}[h]
  \centering
  \caption{Model energy during the MCMC simulations}\label{fig_1}
  \includegraphics[width=0.8\textwidth]{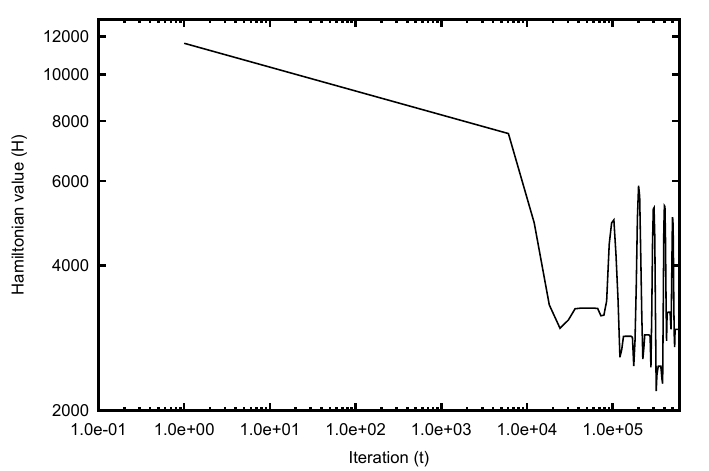}
\end{figure}
The energy variation over the MCMC iterations reported in Figure~\ref{fig_1} shows a progressive and rather rapid decrease in energy during the early iterations of the simulations, 
with the Hamiltonian dropping from an initial high value above 11,000. This decrease is 
followed by an evident stationary process where the energy fluctuates within a narrow range between 2,000 and 3,000. This behavior indicates 
convergence towards optimal solutions, corresponding to a
minimum of the energy ($\Delta H \rightarrow 0$).
During the MCMC simulations, the minimization of $\Delta H$ may be interpreted as being a stochastic gradient descent method applied to the Hamiltonian landscape. The reported stabilization of $\Delta H$ suggests that the 
system has reached a steady state in which the overall energy becomes approximately 
constant over time and concomitantly corresponding to minimum-energy configurations.
The following sections report the results obtained from this simulation process, 
starting from the final energy configuration reached through the annealing procedure.

\subsection{Observed and estimated distributions}

The classification accuracy is calculated by comparing the observed 
configuration ($\textbf{s}_{\text{ref}}$) with the predicted one 
($\textbf{s}_{\text{pre}}$), which is evaluated by summarizing all the 
configurations sampled from the stationary distribution of the 
Ising model; the resulting value is equal to $93.89\%$.
\begin{table}[h]
\centering
\caption{Mismatch matrix}\label{tab_3}
\begin{tabular}{lcc}
\toprule
reference/predicted & No Central Hub (-1) & Central Hub (+1) \\
\midrule
No Central Hub (-1) & 566 & 0 \\
Central Hub (+1)     & 59  & 341 \\
\bottomrule
\end{tabular}
\end{table}
In order to further evaluate the similarity between the observed 
territorial configuration and the probabilities estimated by the 
MCMC simulations, we computed the \emph{Jensen-Shannon Divergence} 
(JSD) between the binary vector derived from the observed configuration 
and the estimated probability distribution $\hat{P}$. The observed 
data (with $1$ indicating a central hub and $0$ otherwise), and $\hat{P}$ 
the vector of estimated probabilities for each municipality. The JSD 
between $P$ and $\hat{P}$ was calculated as:
$\mathrm{JSD}(P, \hat{P}) = 0.0766$
This low divergence value indicates a strong agreement between the 
observed and estimated spatial configurations, confirming the ability 
of the Ising model and the MCMC simulation process to capture the 
underlying territorial structure.

\subsection{Energy and likelihood of the generated configurations}

The energy of the reference configuration was:
$H_{ref} = 11552.59$. 
In order to assess the quality of the configurations 
generated by the model in relation to the observed 
configuration, the ratio between energy $H$ of $25000$ 
sampled configurations and $H_{ref}$ was evaluated. 
Similarly, the log-likelihood ratio of the same was computed 
as described in Equation~\ref{eq:ll_ratio}. 
Results are reported in Table~\ref{tab_4}. 
\begin{table}[h]
\caption{Energy and log-likelihood ratio}\label{tab_4}
\centering
\begin{tabular}{l c c}
\toprule
\textit{Statistic} & \textit{Energy ratio} $H/H_{\text{ref}}$ & \textit{Log-likelihood ratio} $\log(L/L_{\text{ref}})$ \\
\midrule
Min.      & $0.1534$ & $8.7056$ \\
1st Qu.   & $0.2603$ & $8.9044$ \\
Median    & $0.3122$ & $8.9793$ \\
Mean      & $0.3106$ & $8.9817$ \\
3rd Qu.   & $0.3618$ & $9.0530$ \\
Max.      & $0.4780$ & $9.1872$ \\
\bottomrule
\end{tabular}
\end{table}
An alternative bootstrap analysis was also carried out in order to 
investigate the variability of both the log-likelihood ratio and the 
system energy based on the configurations generated by the model. 
A total of $R=200$ bootstrap replications were executed, each consisting 
of a sample of $M=1000$ configurations drawn with replacement from the original set of 
$N$ configurations. For each resample, the mean log-likelihood ratio 
and the mean energy were computed separately. 
Parallel computing was required by allocating $6$ cores of the available 
machine resources. The total computation time was approximately $3$ minutes for each 
bootstrap procedure.
The estimated mean log-likelihood ratio and its $95\%$ confidence interval as well 
as those pertaining to the energy ratio are summarized in 
Table~\ref{tab:bootstrap_summary}.
\begin{table}[h]
\caption{Confidence intervals by using bootstrap}
\label{tab:bootstrap_summary}
\centering
\begin{tabular}{l c c}
\toprule
\textit{Quantity} & \textit{Mean} & \textit{95\% Confidence Interval} \\
\midrule
Log-likelihood ratio & $7964.198$ & $[7920.63,\ 8000.07]$ \\
Energy ratio        & $3591.01$  & $[3547.908,\ 3633.898]$ \\
\bottomrule
\end{tabular}
\end{table}

\subsection{Coverage and adaptivity of the prediction intervals}\label{sec:CP_res}

\begin{figure}[h]
  \centering
  \caption{Model Uncertainty map of Central Italy}\label{fig_2}
  \includegraphics[width=0.8\textwidth]{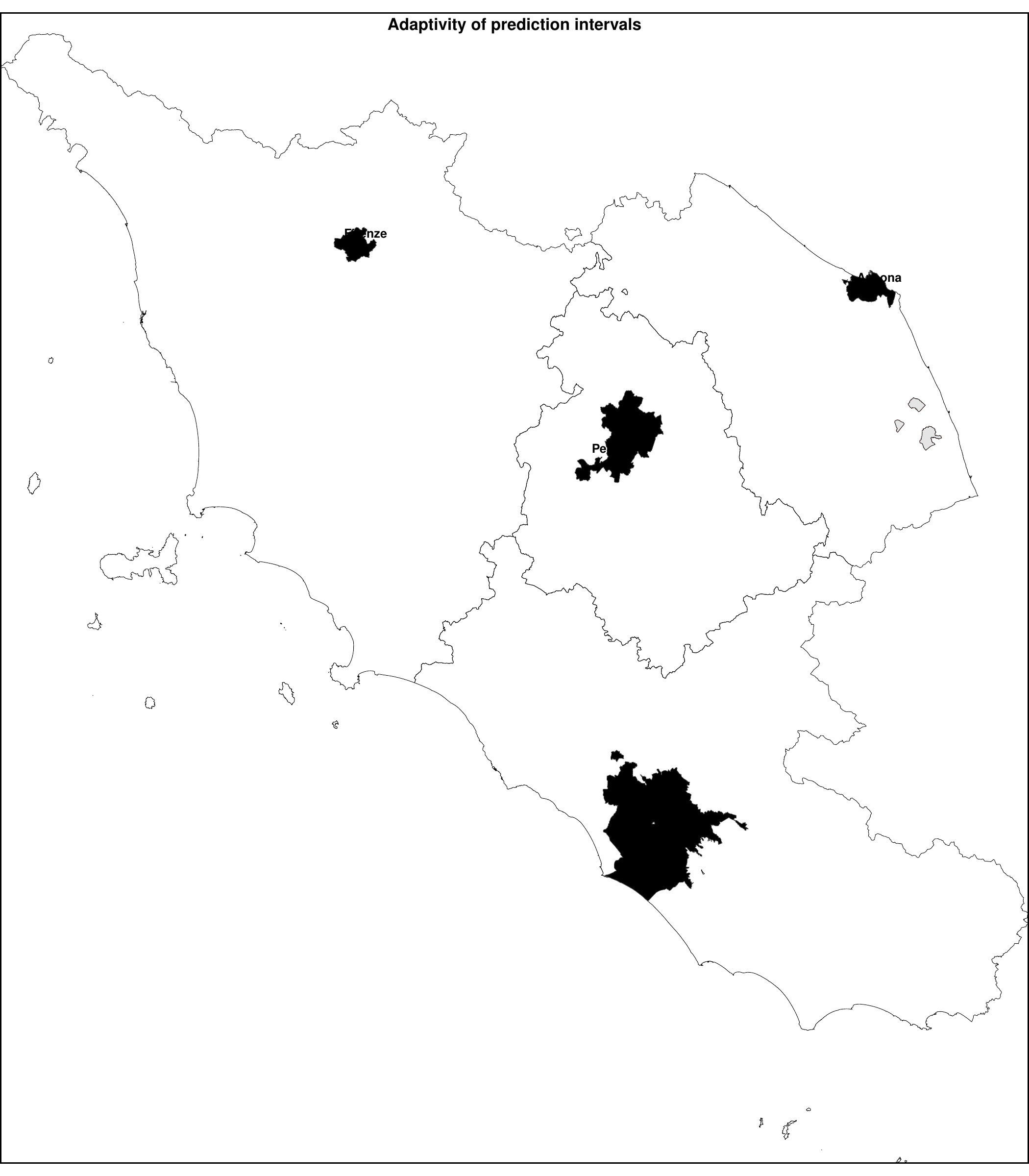}
\end{figure}

In order to evaluate the prediction intervals $K=20000$ samples 
of $N=300$ configurations of spins were generated as is described 
in Section~\ref{sec4_5}. 
Due to the fact that the Ising model estimates the probability 
of being a central hub for each municipality, prediction intervals 
have to be equal to $[0, 1]$ at maximum so that the lower $L$ and the 
upper $U$ bounds of the intervals are required to be corrected as  
follows: $L^{*}=\text{max}(0, L)$ and $U^{*}=\text{min}(U, 1)$ respectively.
As a consequence, the mean interval width (\emph{MIW}) and its 
normalized  version \emph{RIW} coincide. 
The empirical coverage of the prediction intervals results equal to 
$96.99\%$ in accordance with a confidence level equal to $95\%$. 
The number of remaining municipalities with an out-of-interval estimated 
probability is equal to $29$, namely  
a subset of the $59$ mismatches reported in Table~\ref{tab_3}. 
The measurement of the adaptivity reveals that the $94.02\%$ of 
prediction intervals have zero width ($L^{*}=U^{*}$) 
while the percentage of intervals which reveal full uncertainty 
($[L^{*}, U^{*}] = [0, 1]$) is equal to $4.80\%$ and a value of 
$1.17\%$ with intermediate width ($0 < [L^{*}, U^{*}] < [0, 1]$). 
The subset of the remaining $30$ misclassified units 
pertaining to cases in which all the prediction intervals range 
from $0$ to $1$ even though the estimated probability falls into 
the interval. 
In virtue of these results the average adaptivity of the prediction 
intervals $MIW$ is equal to $0.05236$ which suggests an overall 
high accuracy of the model in generating optimal configurations of 
the municipalities. 
This analysis of the adaptivity of the prediction intervals, 
since it is performed only on the subset of cases in which the 
predicted probability falls within the intervals, reveals that 
there are further cases in which the model is uncertain.
To be more precise, $15$ municipalities belonging to the set of 
intervals with the desired coverage were identified as having a 
prediction interval with maximum width indicating maximum model 
uncertainty while $11$ cases with a smaller non-zero interval 
width still indicate uncertainty in the probability estimates of 
those municipalities. 
The uncertainty map illustrates the adaptivity on the 
territory being under examination as is reported in 
Figure~\ref{fig_2}.   
The map focuses exclusively on municipalities whose predicted 
probability of being a central hub falls into the corresponding 
prediction interval.
The figure displays municipalities with non-zero prediction 
interval width in order to indicate the less reliable model results only. 
These latter are possibly due to local data variability, structural model limitations as well 
as insufficient covariate information, impairing interpretation of 
the results so that they require potential methodological refinement.
Municipalities which have full uncertainty 
(i.e., prediction interval width equal to $[0,1]$) are marked in black, 
highlighting maximum model uncertainty while municipalities 
related to lower non-zero values of uncertainty are marked 
in grey. 
Municipalities highlighted on the map do not fully match all the 
misclassified ones reported in Table~\ref{tab_3}, implying that 
there are also units correctly classified even though the model 
is uncertain. Among the $937$ municipalities having a coverage of 
$95\%$, neither the $881$ municipalities pertaining to intervals 
with zero width are misclassified nor the $11$ municipalities with 
non-zero width less than $1$. A residual set of $45$ municipalities 
pertaining to intervals with maximum width are divided into $15$ 
correctly classified units and $30$ misclassified units. 
Results also illustrate that there are $29$ municipalities 
have a probability of being a central hub outside their prediction 
intervals. 

\subsection{Analysis of the socio-economic aspects}

In order to avoid confusion in the reading of this section it is 
important to clarify that unlike the uncertainty map in Section~\ref{sec:CP_res}, 
the analysis in the following includes all observed cases, regardless 
of whether the predicted probability falls into its prediction interval. 
The $59$ misclassified municipalities therefore include both in-interval 
and out-of-interval cases. 
The average values of the input composite indices across three 
groups of municipalities, classified according to the concordance 
between their observed and predicted attractiveness status 
are summarized in Table~\ref{tab_mpi:1}.   .
\begin{table}[h]
\centering
\caption{Average MPI values by classification group}
\label{tab_mpi:1}
\begin{tabular}{lrrrrrrr}
\toprule
Group & n & MPI1 & MPI2 & MPI3 & MPI4 & MPI5 & MPI6 \\
\midrule
No Central Hub-No Central Hub & 566 & 95.92 & 99.90 & 97.83 & 98.30 & 98.46 & 95.92 \\
Central Hub-No Central Hub    &  59 & 98.29 & 101.87 & 101.44 & 100.89 & 97.97 & 98.29 \\
Central Hub-Central Hub       & 341 & 103.55 & 104.64 & 102.45 & 102.87 & 102.69 & 103.55 \\
\bottomrule
\end{tabular}
\end{table}
The average values of the composite indices in accordance with the territorial 
variables being considered in order to create the network of the model are 
reported in Table~\ref{tab_mpi:ALT}--Table~\ref{tab_mpi:DEGURB}.
\begin{table}[h]
\centering
\caption{Average MPI values by group and altitude level}
\label{tab_mpi:ALT}
\begin{tabular}{llrrrrrrr}
\toprule
Group & Altitude & n & MPI1 & MPI2 & MPI3 & MPI4 & MPI5 & MPI6 \\
\midrule
No Central Hub - No Central Hub & 1 & 151 & 100.78 & 100.90 & 97.17 & 97.90 & 99.84 & 100.78 \\
No Central Hub - No Central Hub & 2 & 291 & 96.47  & 100.68 & 98.61 & 99.08 & 98.91 & 96.47 \\
No Central Hub - No Central Hub & 3 & 124 & 88.69  & 96.85  & 96.83 & 96.96 & 95.73 & 88.69 \\
Central Hub - No Central Hub    & 2 &  59 & 98.29  & 101.87 & 101.44 & 100.89 & 97.97 & 98.29 \\
Central Hub - Central Hub       & 1 & 268 & 104.00 & 104.64 & 102.70 & 103.23 & 102.72 & 104.00 \\
Central Hub - Central Hub       & 2 &  65 & 102.41 & 104.79 & 101.82 & 101.88 & 102.51 & 102.41 \\
Central Hub - Central Hub       & 3 &   8 & 97.62  & 103.20 & 99.17  & 98.63  & 103.47 & 97.62 \\
\bottomrule
\end{tabular}
\end{table}
\begin{table}[h]
\centering
\caption{Average MPI values by group and surface area class}
\label{tab_mpi:SUP}
\begin{tabular}{llrrrrrrr}
\toprule
Group & Surface Area & n & MPI1 & MPI2 & MPI3 & MPI4 & MPI5 & MPI6 \\
\midrule
No Central Hub - No Central Hub & 1 &  93 & 94.78  & 98.47  & 97.39 & 97.06 & 94.90 & 94.78 \\
No Central Hub - No Central Hub & 2 & 393 & 96.15  & 99.91  & 97.66 & 98.25 & 98.50 & 96.15 \\
No Central Hub - No Central Hub & 3 &  80 & 96.08  & 101.53 & 99.22 & 100.00 & 102.44 & 96.08 \\
Central Hub - No Central Hub    & 2 &  59 & 98.29  & 101.87 & 101.44 & 100.89 & 97.97 & 98.29 \\
Central Hub - Central Hub       & 1 &  49 & 103.34 & 103.82 & 101.83 & 102.74 & 98.70 & 103.34 \\
Central Hub - Central Hub       & 2 & 220 & 104.28 & 103.99 & 102.37 & 102.99 & 101.72 & 104.28 \\
Central Hub - Central Hub       & 3 &  72 & 101.46 & 107.15 & 103.10 & 102.57 & 108.40 & 101.46 \\
\bottomrule
\end{tabular}
\end{table}
\begin{table}[h]
\centering
\caption{Average MPI values by group and population size class}
\label{tab_mpi:POP}
\begin{tabular}{llrrrrrrr}
\toprule
Group & Population Class & n & MPI1 & MPI2 & MPI3 & MPI4 & MPI5 & MPI6 \\
\midrule
No Central Hub - No Central Hub & 1 & 447 & 93.96  & 99.22  & 97.57 & 98.13 & 97.55 & 93.96 \\
No Central Hub - No Central Hub & 2 & 116 & 103.06 & 102.45 & 98.85 & 98.94 & 101.66 & 103.06 \\
No Central Hub - No Central Hub & 3 &   3 & 110.57 & 102.74 & 98.31 & 98.51 & 110.45 & 110.57 \\
Central Hub - No Central Hub    & 1 &  59 & 98.29  & 101.87 & 101.44 & 100.89 & 97.97 & 98.29 \\
Central Hub - Central Hub       & 1 &  90 & 101.92 & 103.02 & 101.20 & 102.28 & 98.86 & 101.92 \\
Central Hub - Central Hub       & 2 & 226 & 104.34 & 104.84 & 102.88 & 103.31 & 103.15 & 104.34 \\
Central Hub - Central Hub       & 3 &  25 & 102.27 & 108.61 & 103.08 & 100.97 & 112.41 & 102.27 \\
\bottomrule
\end{tabular}
\end{table}
\begin{table}[h]
\centering
\caption{Average MPI values by group and coastal municipality status}
\label{tab_mpi:CLITO}
\begin{tabular}{llrrrrrrr}
\toprule
Group & Coastal (CLITO) & n & MPI1 & MPI2 & MPI3 & MPI4 & MPI5 & MPI6 \\
\midrule
No Central Hub - No Central Hub & 0 & 540 & 95.80 & 99.77 & 98.09 & 98.49 & 98.14 & 95.80 \\
No Central Hub - No Central Hub & 1 &  26 & 98.24 & 102.60 & 92.49 & 94.47 & 105.24 & 98.24 \\
Central Hub - No Central Hub    & 0 &  59 & 98.29 & 101.87 & 101.44 & 100.89 & 97.97 & 98.29 \\
Central Hub - Central Hub       & 0 & 288 & 104.06 & 104.09 & 102.74 & 103.32 & 102.02 & 104.06 \\
Central Hub - Central Hub       & 1 &  53 & 100.75 & 107.63 & 100.85 & 100.41 & 106.36 & 100.75 \\
\bottomrule
\end{tabular}
\end{table}
\begin{table}[h]
\centering
\caption{Average MPI values by group and degree of urbanization}
\label{tab_mpi:DEGURB}
\begin{tabular}{llrrrrrrr}
\toprule
Group & DEGURB & n & MPI1 & MPI2 & MPI3 & MPI4 & MPI5 & MPI6 \\
\midrule
No Central Hub - No Central Hub & 2 &  75 & 105.75 & 102.91 & 97.87 & 98.08 & 102.16 & 105.75 \\
No Central Hub - No Central Hub & 3 & 491 &  94.41 &  99.44 & 97.83 & 98.33 &  97.90 &  94.41 \\
Central Hub - No Central Hub    & 3 &  59 &  98.29 & 101.87 & 101.44 & 100.89 &  97.97 &  98.29 \\
Central Hub - Central Hub       & 1 &  12 & 101.13 & 109.53 & 103.02 & 100.39 & 114.85 & 101.13 \\
Central Hub - Central Hub       & 2 & 202 & 104.70 & 104.93 & 102.63 & 102.97 & 103.91 & 104.70 \\
Central Hub - Central Hub       & 3 & 127 & 101.94 & 103.70 & 102.11 & 102.94 &  99.61 & 101.94 \\
\bottomrule
\end{tabular}
\end{table}
This stratified analysis reveals that misclassified municipalities 
often exhibit intermediate or ambiguous socio-economic profiles, 
confirming the capacity of the model in highlighting ambiguous 
patterns in the observed classification.
The results indicate that the composite indices reveal a clear difference in socio-economic
profiles between municipalities which are classified as being central 
hubs and peripheral areas. The former consistently exhibit higher values on
indices related to economic well-being, employment levels, and territorial
attractiveness, whereas the latter tend to score lower on these
dimensions but higher on indices reflecting demographic aging and population
stability. 
This divergence underscores the fact that complex, multidimensional
processes-ranging from labor market dynamics to population mobility-coalesce to
form the territorial patterns observed within the input data. The synthesis of diverse base indicators 
into coherent composite measures captures not only specific socio-economic aspects but also 
their relationship within a territorial framework, thereby highlighting the 
requirement for models capable of integrating structural territorial information and 
socio-economic latent dynamics, such as the Ising-based approach 
proposed.")n a policy-making perspective, the ability to detect 
ambiguous or transitional municipalities enables a more granular 
allocation of resources and planning strategies, promoting equitable 
development and reducing structural disparities across regions.

\subsection{Benchmark models for comparison}\label{sec:benchmark}

In order to evaluate the performance of the proposed model, three benchmark models 
were estimated using the same explanatory variables described in Section 
\ref{sec4_0}, restricted to municipalities in the central macro-region 
of Italy. These models represent standard approaches from statistical 
inference, spatial econometrics, and machine learning.
The \textit{Logistic regression} is a classical baseline for binary classification. 
It models the log-odds of the probability of being a central hub as a linear 
combination of the six composite indicators, without accounting for any spatial 
or structural dependence among municipalities.
The \textit{Spatial autoregressive} model (SAR) explicitly incorporates a structural 
dependency term by including a weighted average of the dependent variable across 
neighbouring units. In this study, the SAR model was implemented using the same 
conceptual similarity matrix adopted in the proposed Ising-based approach, thus 
allowing a structurally consistent comparison.
The \textit{Random Forest} model is a non-parametric ensemble method based on decision 
trees. It allows for non-linear interactions between variables and does not 
assume any form of spatial structure.
The \textit{Spatial Error Model (SEM)} was intentionally excluded from the 
comparison, as its formulation assumes spatial autocorrelation in the residuals 
rather than in the response variable. Since the aim of the present work is to 
explicitly model structural interdependencies, rather than to absorb unobserved 
spatial effects, the SEM is conceptually misaligned with the proposed modeling 
framework.
Model predictions were compared with the observed classification, which 
corresponds to the initial configuration used in the Ising simulations. The 
comparison was carried out in terms of classification accuracy and Jensen--Shannon 
divergence with respect to the observed configuration.
\begin{table}[h]
\centering
\caption{Classification accuracy and Jensen--Shannon divergence with respect 
to the observed configuration for each benchmark model and the proposed Ising 
model.}
\label{tab:benchmark}
\begin{tabular}{lcc}
\toprule
\textit{Model} & \textit{Accuracy} & \textit{Jensen--Shannon Divergence} \\
\midrule
Logistic Regression          & 0.7536 & 0.1767 \\
Spatial Autoregressive (SAR) & 0.7660 & 0.1811 \\
Random Forest                & 0.7402 & 0.1767 \\
Ising model 	& 0.9389 & 0.0766 \\
\bottomrule
\end{tabular}
\end{table}
These benchmark results confirm that the proposed Ising-based model 
achieves a satisfactory balance between structural coherence and 
predictive performance. While the logistic and spatial autoregressive 
models rely on linear assumptions or fixed spatial structures, and 
Random Forest captures non-linearities without accounting for spatial 
dependencies, the Ising model is able to embed both structural 
interactions and explanatory heterogeneity. Despite not being optimized 
for classification accuracy, it attains a competitive performance, supporting 
its suitability for modeling complex territorial dynamics.

\section{Conclusions}\label{sec6}

This study introduces a framework for modeling territorial dynamics 
based on the Ising model, applied to the classification of 
municipalities into central hubs and peripheral areas. The approach 
integrates a spatial interaction network constructed from shared 
structural and geographical characteristics with an external field 
defined by socio-economic composite indices. This formulation allows 
the model to incorporate both local interactions and external 
influences in a unified probabilistic context.
The use of composite indices serves a dual purpose: in addition to 
providing a concise and interpretable representation of complex 
socio-economic phenomena, these indices are a strategy for 
dimensionality reduction concomitantly taking 
multiple interrelated factors into account. This contributes to a more tractable 
modeling framework whilst preserving the multidimensional nature of 
territorial dynamics. 
Model simulations leverage the Markov Chain Monte Carlo method which 
employs a Simulated Annealing variant. 
In order to reduce the computational effort of MCMC simulations, it 
is necessary to adopt a parallel computing approach so that the execution time is 
kept at an acceptable level.  This ensures scalability of the 
proposed approach when applied to more complex large-scale datasets.
This study may represent one of the first applications of a statistical 
mechanics-based model for territorial classification driven by 
structural socio-economic similarities in substitution of geographical 
contiguity. The proposed framework may be extended to other domains in 
Official Statistics in which latent interaction structures emerge from 
demographic and economic indicators. The objective is to search for 
configurations which are energetically more favorable and 
statistically more likely in order to compare them to the 
observed reference. As the Markov chain becomes stationary, 
the algorithm explores configurations which are close to the one 
pertaining to a minimum energy value and providing insight into 
alternative structural equilibria.
The application of Conformal Prediction enhances the proposed framework by 
enabling the construction of adaptive prediction intervals for each 
local classification. These intervals serve as a valuable measure of 
uncertainty varying across the territory in accordance with the strength of local 
external influences. In this context, uncertainty maps are introduced as an 
original analytical tool, capable of revealing territorial sub-areas 
where the model exhibits greater unreliability. These maps 
support a more detailed interpretation of local dynamics.
By combining elements of statistical physics, multivariate analysis, 
and probabilistic inference, the proposed approach offers a flexible 
and extensible methodology for the analysis of spatial classifications. It is suitable for 
a wide range of territorial applications which include: multidimensional policy 
analysis, identification of structurally ambiguous areas with potential 
extensions to spatial-temporal systems.
The set-up of a similarity-based graph which connects 
municipalities having analogous territorial profiles is 
unusual when compared to models which rely on traditional spatial 
contiguity. This structural perspective encompasses interactions 
which are more meaningful in the socio-economic context, extending 
the applicability of the Ising model to conceptual networks rather 
than strictly spatial ones.
The empirical results suggest that the proposed Ising-based approach 
captures the latent structure of territorial systems, offering 
a novel perspective for interpreting spatial configurations grounded in 
statistical mechanics. While the model was not primarily designed for 
predictive purposes, its classification performance is competitive when 
benchmarked against standard statistical and machine learning methods. 
This indicates that the proposed methodology not only reproduces observed 
patterns but also embeds a level of robustness that may support its 
application in real-world territorial analyses. 
Future research may explore continuous-valued generalizations of the model or 
extend the analysis to finer spatial scales such as provinces or census tracts, 
allowing for a deeper investigation of local heterogeneities and uncertainty 
sources.

\section*{Declarations}

\begin{itemize}
  \item \textbf{Funding} \\  
    The author declares that this research received no external funding.
	\\
  \item \textbf{Conflict of interest/Competing interests} \\  
    The author declares no conflicts of interest regarding the content of this manuscript.
  \\
   \item \textbf{Ethics approval and consent to participate} \\  
    Not applicable.
    \\ 
  \item \textbf{Consent for publication} \\  
    Not applicable.
   \\  
  \item \textbf{Data availability} \\  
    Not applicable. 
    \\ 
  \item \textbf{Materials availability} \\  
    Not applicable. 
    \\ .
	\item \textbf{Code availability} \\  
	The code supporting the findings 
	of this study is available upon 
	reasonable request. \\
\end{itemize}




\end{document}